# Beyond Single-Dimension Novelty: How Combinations of Theory, Method, and Results-based Novelty Shape Scientific Impact

Yi Zhao[1][0000-0003-1050-7051] Chenggang Yang[2][0009-0000-2984-1733] Yuzhuo Wang[1][0000-0002-2891-7238] Tong Bao[2][0000-0003-1871-8005] Heng Zhang[3][0009-0007-6751-7062] and Chengzhi Zhang[2*][0000-0001-9522-2914]

[1] School of Management, Anhui University, Hefei 230039, China
{yizhao93,wangyuzhuo}@ahu.edu.cn
[2] Department of Information Management, Nanjing University of Science and Technology, Nanjing, 210094, China
{ichigo,tbao,zhangcz}@njust.edu.cn
[3] School of Information Management, Central China Normal University, Wuhan 430079, China
zh_heng@ccnu.edu.cn

**Abstract.** Scientific novelty drives advances at the research frontier, yet it is also associated with heightened uncertainty and potential resistance from incumbent paradigms, leading to complex patterns of scientific impact. Prior studies have primarily examined the relationship between a single dimension of novelty—such as theoretical, methodological, or results-based novelty—and scientific impact. However, because scientific novelty is inherently multidimensional, focusing on isolated dimensions may obscure how different types of novelty jointly shape impact. Consequently, we know little about how combinations of novelty types influence scientific impact. To this end, we draw on a dataset of 15,322 articles published in *Nature Communications*. Using the DeepSeek-V3 model, we classify articles into three novelty dimensions based on the content of their Introduction sections: theoretical novelty, methodological novelty, and results-based novelty. These dimensions may coexist within the same article, forming distinct novelty configurations. Scientific impact is measured using five-year citation counts and indicators of whether an article belongs to the top 1% or top 10% highly cited papers. Descriptive results indicate that results-based novelty alone and the simultaneous presence of all three novelty types are the dominant configurations in the sample. Regression results further show that articles with results-based novelty only receive significantly more citations and are more likely to rank among the top 1% and top 10% highly cited papers than articles exhibiting all three novelty types. These findings advance our understanding of how multidimensional novelty configurations shape knowledge diffusion.

---

* Corresponding author.

# 1   Introduction

Scientific novelty is widely regarded as a central driver of advances at the scientific frontier and of technological progress. Although novel research holds the potential to generate substantial scientific impact, it is also inherently associated with considerable uncertainty [1]. Clarifying the relationship between the novelty of scholarly work and its subsequent scientific impact has therefore attracted sustained attention from the scientific community. Agovino et al. (2017) argue that novelty constitutes a precondition for generating downstream impact [2]. Several empirical studies suggest that novel research can stimulate follow-up investigations, thereby producing broader and more enduring influence [3]. At the same time, highly novel research may challenge established paradigms and thus encounter resistance from prevailing scientific communities [4]. Empirical evidence indicates that novel contributions are more likely to be rejected during peer review, published in lower-tier journals, or receive longer handling time [5]. Other scholars propose that moderate levels of novelty are more likely to attract widespread attention, implying an inverted U-shaped relationship between novelty and scientific impact[6].

Existing research suggests that novelty is not a unidimensional construct but encompasses multiple dimensions, including theoretical, methodological, and results-based novelty. For example, Wu et al. (2025) identify methodological novelty as one of the most prevalent forms of novelty in scholarly articles [7]. In practice, a single article may simultaneously exhibit theoretical, methodological, and empirical novelty, and its innovative value may not be a simple linear aggregation of these elements. Rather, novelty may assume a configurational structure characterized by specific combinations of dimensions. Despite this recognition, the distributional patterns of different novelty configurations across scientific domains remain insufficiently examined. Moreover, distinct types of novelty may follow heterogeneous diffusion trajectories and attract attention through different evaluative and cognitive mechanisms[8]. For instance, Leahey et al. (2023) show that new methods tend to be more disruptive, whereas new theories are comparatively less so [9]. This implied that when multiple novelty dimensions coexist within the same article, their combined structure may therefore exert interactive—rather than independent—effects on scientific impact. These considerations suggest that analyses focusing on a single dimension of novelty may offer an incomplete account of how innovation translates into scientific influence. A systematic examination of the distribution of novelty configurations and their differential impact effects can deepen our understanding of the mechanisms underlying knowledge production and dissemination, while also informing the refinement of science evaluation systems. To this end, we propose the following research question：

*RQ1*: Which combinations of theory, method, and result novelty are more prevalent in scholarly articles?

*RQ2*: Do different combinations of novelty types exert significantly different effects on the scientific impact of scholarly articles?

## 2   Methodology

### 2.1   Dataset

The dataset analyzed in this study comprises research articles published in *Nature Communications*, a leading multidisciplinary journal. We examine 15,327 articles published between 2016 and 2021. To augment the publication records with comprehensive bibliographic and citation information, we linked these articles to SciSciNet [10], an openly accessible, large-scale scholarly data lake containing metadata for over 134 million scientific publications and millions of inter-entity linkages (e.g., citations, grants, and patents). SciSciNet has become an important data lake for bibliometric and science-of-science research due to its scale and integrative structure [11]. After excluding retracted articles, 15,322 publications were successfully matched to records in SciSciNet and retained for subsequent analyses.

SciSciNet derives its core publication metadata from the Microsoft Academic Graph (MAG), which adopts a six-level hierarchical disciplinary classification system (L0–L5). The L0 level represents the broadest categorization and comprises 19 top-level disciplines. Given that our dataset contains 15,322 articles, some L0 disciplines are represented by only a small number of publications, potentially limiting statistical reliability. To improve analytical robustness, we mapped the 19 MAG L0 disciplines to the Scopus All Science Journal Classification (ASJC) scheme and consolidated them into four broad domains: Physical Sciences, Health Sciences, Social Sciences, and Life Sciences. Of the 15,322 articles, 7,284 (47.54%) are classified as Physical Sciences, 6,554 (42.78%) as Life Sciences, 1,068 (6.97%) as Health Sciences, and 416 (2.72%) as Social Sciences.

### 2.2   Classification of novelty types in scholarly articles

Following the practice of our previous work [12], this study extracts sentences describing novel contributions from the introduction sections of research articles and classifies them to characterize the types of novelty embodied in scholarly work. Drawing on the framework proposed by Leahey et al. (2023), we distinguish three types of novelty: theoretical novelty, methodological novelty, and results-based novelty[9]. A single article may contain multiple sentences articulating novel contributions, and these sentences may correspond to different types of novelty.

To automate the extraction and classification of novelty statements, we employed the DeepSeek-V3 model with carefully designed prompts (see Table A.1 in the Appendix). As journal articles typically undergo rigorous peer review, their content provides a reliable basis for identifying and categorizing claims of novel contribution. Operationally, an article is coded as exhibiting a given type of novelty if at least one sentence corresponding to that novelty type is identified within its text.

To evaluate the classification performance of the DeepSeek-V3 model, we manually annotated a random sample of 50 articles to construct a gold-standard dataset for subsequent evaluation. The evaluation results indicate that the model effectively identifies sentences corresponding to the predefined novelty types, achieving an $F_1$ score of 0.77. This performance indicates a satisfactory level of reliability in identifying novelty types in scholarly articles.

### 2.3    Dependent, independent, and control variables

**(1)  Dependent variable**

Citation counts are widely used as indicators of the scientific impact of scholarly publications [11, 13]. Following prior research [14], we used two types of forward citations: five-year citation (continuous variable) and hit papers (binary variable). The former is defined as the number of citations a paper receives within five years of publication. The latter identifies papers ranked in the top 1% and top 10% of the citation distribution within the same field and publication year. Using both measures allow us to capture overall impact as well as the likelihood of producing highly influential papers.

**(2)  Independent variable**

We classify scientific novelty into three types: theoretical novelty, methodological novelty, and results-based novelty. Because an article may exhibit one or more of these types, eight configurations are theoretically possible, including the absence of all three. In our sample, however, no article is characterized by theoretical novelty alone. Consequently, seven distinct novelty configurations are observed and retained for subsequent analyses.

Given the multidimensional nature of novelty, we model these configurations as a categorical variable. The seven observed combinations are represented by a set of mutually exclusive dummy variables in the regression models, with the configuration exhibiting all three types of novelty (theoretical, methodological, and results-based) serving as the reference category.

**(3)  Control variables**

Consistent with prior research, citation counts may be influenced by a range of factors that can be broadly classified into author-level and paper-level characteristics [11]. To mitigate potential confounding effects, we incorporate both sets of variables as controls in the regression models. Table 1 summarizes the definitions of the control variables.

**Table 1.** List of control variables.

| Control variables | Description |
|---|---|
| Team size | The total number of authors listed for the focal paper. |
| Team reputation | The mean H-index of all co-authors of a given article. |
| International collaboration | A team is classified as an international collaboration if it includes members from at least two distinct countries; such teams |
| | are coded as 1. Teams with members from only one country are coded as 0. |
| Reference count | The total number of references cited by the focal paper. |
| Publication year | The year in which the focal paper was published. The publication year is included as a fixed effect to control for time-related factors that may influence citation patterns. |
| Research domains | The research domain of the focal paper. The research domain is included as a fixed effect to control for unobserved factors that may arise from varying citation patterns across different domains. |

### 2.4 Regression models

Multivariable regression was employed to examine the relationship between novelty type combinations and scientific impact. The empirical model is specified as follows:

$$C_p = \beta_0 + \sum_{i=1}^{6} \gamma_i \cdot NovComb_{i,p} + \beta_1 X_p + D_p + \varepsilon_p \qquad (1)$$

Where $C_p$ represents the scientific impact of paper $p$. In the main analysis, scientific impact is measured using 5-year citation counts and an indicator for hit papers. $X_p$ denotes a vector of paper-level control variables that may influence impact, as listed in Table 1. $NovComb_{i,p}$ is a set of mutually exclusive dummy variables representing the six observed novelty type combinations. $D_p$ represents fixed effects for publication year and research field to account for unobserved heterogeneity, and $\varepsilon_p$ is the error term. Since 5-year citations are count data, we estimate the corresponding models using negative binomial regression. For the hit paper outcome, which is binary, we adopt logistic regression.

## 3 Results

### 3.1 The distribution of scientific novelty combinations

To address RQ1, we employ the DeepSeek-V3 model to classify novelty into three dimensions: theoretical (T), methodological (M), and results-based (R), and identify their observed configurations. As shown in Table 2, the distribution of novelty configurations is highly skewed. The majority of articles exhibit results-based novelty only (R), accounting for 61.08% of the sample. The second most prevalent configuration is the simultaneous presence of all three types T, M, and R, representing 37.29%. The remaining configurations are rare, each comprising less than 1% of the sample. No article in the dataset is characterized by theoretical novelty alone. Because this study focuses on how combinations of novelty shape scientific impact, articles that do not exhibit any of the three novelty types are excluded from subsequent analyses. In addition, the configuration representing methodological novelty only (M) contains a single observation. Given its extremely small cell size and the potential implications for statistical reliability, this observation is also excluded from the regression analyses.

Table 2. Distribution of novelty combination types

| Novelty combinations | Label | # articles | Share (%) |
|---|---|---|---|
| Results-based novelty only | R | 9,358 | 61.08% |
| Theoretical + Methodological + Results-based novelty | T+M+R | 5,714 | 37.29% |
| Theoretical + Methodological novelty | T+M | 85 | 0.55% |
| No identifiable novelty | None | 74 | 0.48% |
| Theoretical + Results-based novelty | T+R | 49 | 0.32% |
| Methodological + Results-based novelty | M+R | 41 | 0.27% |
| Methodological novelty only | M | 1 | 0.01% |

**Notes**: T = Theoretical novelty; M = Methodological novelty; R = Results-based novelty.

### 3.2   Association between novelty combinations and scientific impact

To answer RQ2, we conducted a series of multivariable regressions to examine the relationship between combinations of novelty types and scientific impact. Table 3 presents these regression results. Model (1) reports negative binomial estimates for five-year citation counts, whereas Models (2) and (3) provide logit estimates for the likelihood of reaching the top 1% and top 10% of the citation distribution, respectively. The reference group is the full novelty type (T+M+R). In model (1), we find that articles with results-based novelty only (R) receive significantly more citations than the reference group ($\beta = 0.176, p < 0.01$). A similar and even stronger positive association is observed for the combination of methodological and results-based novelty (M+R, $\beta = 0.349, p < 0.01$). In contrast, articles combining theoretical and methodological novelty (T+M) receive significantly fewer citations ($\beta = -0.183, p < 0.01$), while the T+R configuration does not differ statistically from the full novelty type. The logit estimates in Models (2) and (3) reveal a similar pattern. Articles with results-based novelty only (R) are significantly more likely to become top 1% ($\beta = 0.359, p < 0.01$) and top 10% ($\beta = 0.275, p < 0.01$) most cited papers. The corresponding odds ratios ($e^{0.359} \approx 1.43$ and $e^{0.275} \approx 1.32$, respectively) suggest substantial performance advantages over T+M+R papers. In addition, Model (2) indicates that the M+R combination is associated with a higher likelihood of producing top 1% cited papers ($\beta = 0.685, p < 0.1$), corresponding to nearly twice the odds ($e^{0.685} \approx 1.98$) compared to the reference group.

Among the control variables, team size and team reputation are positively associated with scientific impact, suggesting that larger and more reputable teams tend to achieve greater visibility. In contrast, international collaboration is negatively associated with citation-based outcomes. Importantly, these patterns remain robust across different measures of impact.

Table 3. Estimated effect of different novelty combinations on scientific impact

| Variable | (1) 5-year citations (NBR) | (2) Top 1% hit paper (logit) | (3) Top 10% hit paper (logit) |
|---|---|---|---|
| R | 0.176*** | 0.359*** | 0.275*** |
|  | (0.024) | (0.046) | (0.044) |
| M+R | 0.349** | 0.685* | 0.386 |
|  | (0.176) | (0.374) | (0.514) |
| T+R | 0.012 | -0.435 | 0.144 |
|  | (0.092) | (0.503) | (0.368) |
| T+M | -0.183** | -0.621 | 0.125 |
|  | (0.075) | (0.404) | (0.276) |
| International collaboration | -0.169*** | -0.332*** | -0.152*** |
|  | (0.024) | (0.046) | (0.048) |
| Team size | 0.218*** | 0.414*** | 0.418*** |
|  | (0.020) | (0.035) | (0.038) |
| Reference count | 0.061 | 0.044 | 0.610*** |
|  | (0.047) | (0.072) | (0.069) |
| Team reputation | 0.012*** | 0.025*** | 0.022*** |
|  | (0.001) | (0.001) | (0.002) |
| _cons | 3.616*** | -2.546*** | 0.007 |
|  | (0.277) | (0.606) | (1.066) |
| lnalpha | -0.483*** |  |  |
|  | (0.025) |  |  |
| Publication year fixed effect | YES | YES | YES |
| Research domain fixed effect | YES | YES | YES |
| Observations | 15078 | 15078 | 15078 |
| Pseudo R2 | 0.0157 | 0.0724 | 0.0532 |

## 4    Discussion and Conclusions

In this study, we examine the relationship between various novelty configurations and scientific impact using a sample of 15,322 articles published in Nature Communications. We first employed a large language model to classify scientific novelty into three dimensions, namely theoretical novelty, methodological novelty, and results-based novelty, thereby identifying seven possible configurations in principle. The classification results indicate that articles characterized by results-based novelty only and those exhibiting the simultaneous presence of all three novelty types dominate the sample, jointly accounting for 98.37% of the observations. Regression analyses further reveal that articles featuring results-based novelty alone receive significantly more citations and are more likely to rank among the top 1% and top 10% of highly cited papers compared to articles combining all three novelty types. These findings suggest that citation rewards tend to concentrate on research emphasizing results-based contributions rather than on work integrating multiple dimensions of novelty. This pattern raises concerns that widely used citation-based measures may undervalue certain forms of multidimensional novelty. Consistent with this interpretation, prior research suggests that novel knowledge often encounters resistance from incumbent paradigms and established interests [4].

This study has several limitations. First, the dataset is drawn exclusively from *Nature Communications*. Although this journal covers a wide range of disciplines, caution is warranted when generalizing the findings to other publication outlets, disciplinary contexts, or larger and more heterogeneous samples. Future research could extend the analysis to broader datasets to assess the robustness and external validity of the results. Second, while the classification performance of the novelty aspects is acceptable, misclassification errors may still occur. Such measurement inaccuracies could introduce bias into the estimated relationships and affect the precision of the empirical results. Further methodological refinements and validation efforts would help improve the reliability of novelty identification.

**Acknowledgments.** This study was funded by Fundamental Research Funds for the Central Universities (grant number: CCNU25XJ037).

**Disclosure of Interests.** The authors declare no conflict of interest.

## Appendix

**Table A1.** Prompt for classifying types of scientific novelty

| | |
|---|---|
| Instruction | **The persona pattern:**<br>You are a proficient linguist skilled in reading academic articles.<br>**Introduce the target of our task:**<br>You will be given a paragraph in the Introduction section of a publication. Please follow the instructions to label the paragraph in the Introduction section provided by user. In the introduction of a paper, the author mentions the innovations of the entire work, which we define as contribution statements.<br>**Definition of contribution statement:**<br>Contribution statements are sentences or paragraphs in academic papers that clearly highlight the main contributions and innovations of the research work.<br>**Definition of three types of innovation:**<br>These contributions can be categorized into three types:<br>Theoretical Innovation: Refers to breakthroughs in theoretical frameworks, models, or concepts. This can involve new theoretical perspectives, redefinitions of concepts, or extensions of existing theories, which advance the understanding and development of the discipline.<br>Methodological Innovation: Involves improvements or innovations in research methods, techniques, or tools. This can include new experimental designs, data |
| | collection methods, and analytical techniques, making research more efficient and reliable, or enabling the resolution of previously unsolvable problems.<br>Results-based Innovation: Refers to new findings or conclusions obtained from the research. This type of innovation emphasizes the new knowledge or data gained from the research and its potential applications, which can have a significant impact on theory, practice, or policy.<br>**Introduce the details in our task:**<br>Each paper's introduction may contain one or more of these three types of innovations, and each sentence belongs to only one type of innovation. Please read the provided research paper's introduction carefully and extract the original sentences representing these contribution statements, categorizing them into theoretical innovation, methodological innovation, and result innovation. No explanations are needed for the extracted results. |
| Input | Introduction of an article. |
| Output | **theoretical innovation**: [extracted theoretical innovation statement](if none, leave blank);<br>**methodological innovation**: [extracted methodological innovation statement](if none, leave blank);<br>**results-based innovation**: [extracted result innovation statement](if none, leave blank)" |